\documentclass[aps,prl,twocolumn,epsf,epsfig,amsmath,colorlinks=true, superscriptaddress, scitecolor=blue, linkcolor=blue, urlcolor=blue]{revtex4-2}
\usepackage{latexsym}
\usepackage{times}
\usepackage{graphicx}
\usepackage{amsmath}
\usepackage{multirow}
\usepackage{amsthm}
\usepackage{dcolumn}
\usepackage{bm}
\usepackage{ textcomp }
\usepackage{color}
\usepackage{amssymb}
\usepackage{xcolor}
\usepackage{easyReview}
\usepackage{mathdots}
\usepackage{float}
\usepackage{lineno}
\usepackage{todonotes}
\usepackage{array}
\usepackage{orcidlink}
\usepackage{booktabs}
\usepackage{stackengine}
\usepackage{braket}
\newcommand{\beq}{\begin{eqnarray}}
\newcommand{\eeq}{\end{eqnarray}}

\begin{document}

\title{Re-entrant p-wave superconductivity in a Chern Dartboard Insulator}

\author{Yueyang Wu$^{1}$, Rebecca Chan$^{1}$, Taylor L. Hughes$^{1}$ and Philip W. Phillips}

\affiliation{Department of Physics and Anthony J. Leggett Institute of Condensed Matter Theory, University of Illinois at Urbana-Champaign, Urbana, IL 61801, USA}

\date{\today}

\begin{abstract}
We formulate an exact theory for a 2-band superconductor with inter-orbital pairing and band Hatsugai-Kohmoto (HK) interactions. We apply this theory to a Chern dartboard insulator and compute the pair susceptibility and pair correlations analytically.  We find that in the insulating phase previously predicted by BCS theory (small pairing strength $g$, chemical potential $\mu$ in the gap), the HK interaction can induce superconductivity, and possibly a Chern dartboard superconductor, by lifting the doubly-occupied band and making it cross the chemical potential. From this HK-induced superconductor, we notice that in a significant parameter range (HK interaction strength $U = 1$, $\mu = 0.5$), an insulator-superconductor-insulator re-entrant transition exists. We isolate the re-entrance to a conflation between HK interactions and inter-orbital pairings.
\end{abstract}

\maketitle

\section{Introduction}

Traditional non-trivial band topology arises from an invariant~\cite{Haldane, KM, BHZ} that is quantized when integrated over the whole Brillouin zone~\cite{ChernNumber}. However, recently~\cite{CDI}, new types of topological crystalline insulators, dubbed Chern dartboard insulators (CDIs), have a quantized first Chern number upon an integration over just a fraction of the allowable Brillouin zone. In particular, n$^{\rm th}$-order CDIs display such quantization by integration over a sector of $1/2n$ of the full Brillouin zone. This quantization can be protected by $n$ mirror symmetries, and as a consequence, such sub-Brillouin zone (sBZ) quantization lies outside the standard ten-fold way~\cite{Ryu_2010} classification scheme as well as topological quantum chemistry~\cite{TopoQuantChemBarry}. 

Because of the current interest in topological superconductivity~\cite{TopoSC, TopoSC2}, He-3~\cite{he3} being the first example, an obvious question with any new platform exhibiting a non-zero Chern number is, what happens in the presence of pairing? Moreover, we can ask how does a CDI superconduct?  In a recent paper, extending earlier work on proximitizing Chern insulators \cite{qhz2010}, Chan and Hughes~\cite{chan2025cherndartboardsuperconductors} addressed the first question by including a mean-field Cooper pairing term~\cite{BCSTheory} in a variety of CDI models. They showed that the resultant systems can realize Chern dartboard superconductors (CDSC) and have non-trivial superconducting sBZ Chern numbers. Furthermore, their work highlighted that some of the CDSC phases can have quantized responses to strain fields. 

As is typical in studies of band-structure-induced topology, their construction is based on non-interacting electrons coupled with a mean-field, proximity-induced \cite{fu2008}, superconducting pairing. In contrast, with the advent of Moir\'e systems, non-trivial band topology can arise in flat bands in which the interactions dominate~\cite{FlatbandsTopology, MoireMagicAngle, FlatbansMoire}. Consequently, it is timely, and perhaps even experimentally relevant, to consider interaction-induced mechanisms to generate a CDSC. Of course, topology and interactions pose a notoriously difficult combination to tame. However, it has recently been shown that the predictions of the Hubbard/Kane-Mele and BHZ~\cite{1over4is1over2}, or Hubbard/Haldane~\cite{1over4is1over2Haldane} models, coincide with those of the Hatsugai-Kohmoto counterpart. This is particularly advantageous, as the HK model~\cite{HKModel, ExactSolvableModel} is exactly solvable even with the Cooper interaction~\cite{phillips_exact_2020, PhysRevB.105.184509}. Combining sBZ topology with HK is precisely what we pursue in this paper. We redo the Cooper analysis of the CDSC in the presence of HK interactions. This amounts to a 2-band spinful generalization of the previous works. Surprisingly, we find that the HK interaction can induce superconductivity in a CDI, where the traditional BCS theory~\cite{BCSTheory} predicts its absence.  Moreover, an insulator-superconductor-insulator re-entrant phenomenon is observed under certain values of the HK interaction strength and chemical potential.

\section{Model}

\begin{figure}[ht]
    \begin{minipage}[c]{0.49\linewidth}
        \centering
\stackinset{l}{-3pt}{t}{-3pt}{\textbf{(a)}}{%
            \includegraphics[width=\linewidth]{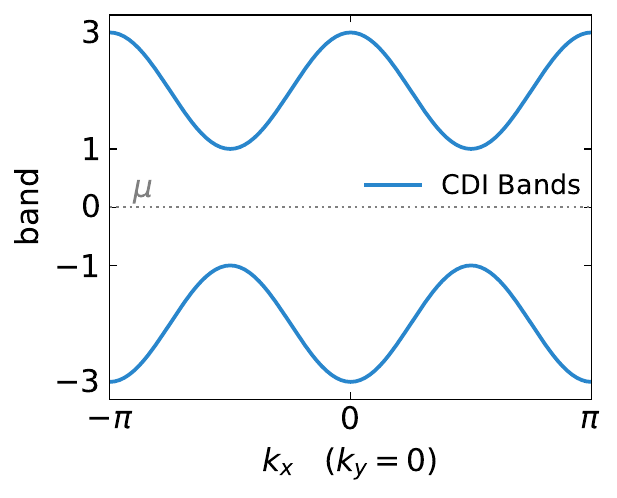}%
        }
    \end{minipage}
    \hfill
    \begin{minipage}[c]{0.49\linewidth}
        \centering
\stackinset{l}{-3pt}{t}{-3pt}{\textbf{(b)}}{%
            \includegraphics[width=\linewidth]{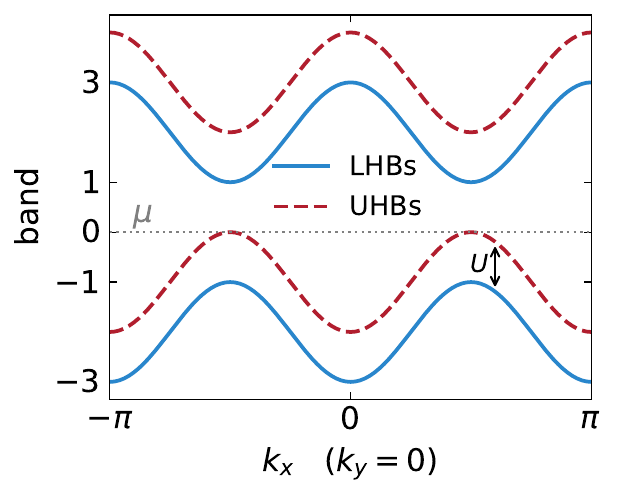}%
        }
    \end{minipage}
    \begin{minipage}[c]{1.0\linewidth}
        \centering
\stackinset{l}{-3pt}{t}{-3pt}{\textbf{(c)}}{%
            \includegraphics[width=\linewidth]{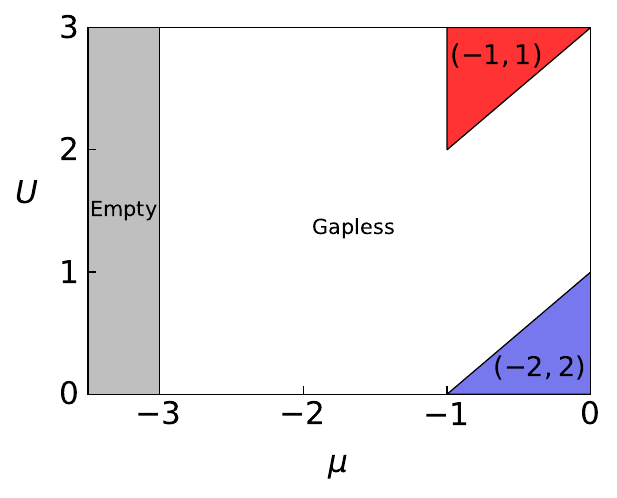}%
        }
    \end{minipage}
    \caption{\textbf{(a)} The bands of our CDI + HK system along the $k_y$ = 0 line for $U$ = 0. We set the chemical potential $\mu = 0$ so that it resides in the band gap. Without a large enough pairing strength $g$, the system insulates. \textbf{(b)} The single-particle Green's function poles of our CDI + HK system along the $k_y$ = 0 line when $U > 0$. The chemical potential $\mu$ is still at 0, but the HK interaction creates the UHBs, thus $\mu$ now touches one of these interaction-induced bands. BCS pairs can now obtain more easily than in the case in \textbf{(a)}, thereby leading to superconductivity. \textbf{(c)} Phase diagram of CDI + HK system at $T=0$. Gapped, incompressible, insulating phases are labeled by their reduced Chern numbers $(N_r^a,N_r^b)$. $N_r^a$ refers to the upper right and lower left quadrants of the BZ, and $N_r^b$ refers to the upper left and lower right quadrants. }
    \label{fig:BandPlot}
\end{figure}

Although superconductivity in the presence of HK has been treated previously~\cite{phillips_exact_2020, PhysRevB.105.184509}, the new ingredients in the CDI problem require multiple bands possessing non-trivial Chern numbers and inter-orbital pairings. Thus, in this work, we will consider a 2-band generalization of previous works. The model Hamiltonian is of the following form
\beq
\label{eq:2BandHKSCHamiltonian}
H=H_{0}+H_{\text{\rm HK}}+H_{\text{\rm pairing}},
\eeq
where $H_{0}$ is the single-particle Hamiltonian in the orbital basis for an $n=2$ Chern dartboard insulator, $H_{\text{HK}}$ is the HK interaction in the band basis, and $H_{\text{pairing}}$ is the superconducting pairing term. Specifically, we have 
\begin{equation}
    H_0 = \sum_k (H_2 - \mu\tau_0) \otimes \sigma_0,
\end{equation}
where
\beq
    H_2 = d_x(\mathbf{k})\tau_x + d_y(\mathbf{k})\tau_y + d_z(\mathbf{k})\tau_z,
\eeq
and $\sigma_\mu (\tau_\nu)$ are Pauli matrices for spin-1/2 (orbitals 1,2). We have $d_x(\mathbf{k}) = \sin(k_x)\sin(2k_y)$, $d_y(\mathbf{k}) = \sin(2k_x)\sin(k_y)$, $d_z(\mathbf{k}) = 1 + \cos(2k_x) + \cos(2k_y)$. After diagonalizing $H_2$,  we obtain $E_k = \sqrt{d_x^2(\mathbf{k}) + d_y^2(\mathbf{k}) + d_z^2(\mathbf{k})}$ which yields the energy dispersions for the 2 bands $\xi_{\mathrm{I},k}^{l} = -E_k - \mu$ and $\xi_{\mathrm{II},k}^{l} = E_k - \mu$. These bands are shown in Fig.~\ref{fig:BandPlot} \textbf{(a)}, and the superscript $l$ denotes the lower Hubbard band, which will be discussed later. The single-particle Hamiltonian $H_0$ supports an $n=2$ CDI phase with protective mirror symmetries $M_{x,y}$ in both the $x$- and $y$-directions, and thus the first Brillouin zone of this model can be divided into 4 sub-Brillouin zones, each with a well-defined sBZ Chern number. Explicitly, for the model parameters we chose, the sBZ Chern numbers take a sign-alternating quadrupolar pattern~\cite{chan2025cherndartboardsuperconductors} with a negative sBZ Chern number in the upper-right quadrant. Since we take the spin-up and -down CDIs to be the same, the sBZ Chern numbers are $\pm2$ at half-filling.



We next include an HK interaction term of the form
\beq
    H_{\text{HK}} = U\sum_k n_{\mathrm{I},k,\uparrow}n_{\mathrm{I},k,\downarrow} + n_{\mathrm{II},k,\uparrow}n_{\mathrm{II},k,\downarrow},
    \label{hkint}
\eeq
where $U$ is the coupling strength. An immediate advantage of the HK interaction over Hubbard~\cite{HubbardModel} is that exact results are possible even for the pair susceptibility. This state of affairs obtains because different momentum sectors in the HK model remain uncoupled as opposed to Hubbard. Since recent work has shown direct links between the HK and Hubbard models generated by a twist in the boundary conditions~\cite{HKConvergeHubbard, MMHK}, we can view the use of the HK model as a simple starting point for a more involved momentum-coupling calculation in future work. 

The phase diagram as a function of $U$ and $\mu$ is shown in Fig.~\ref{fig:BandPlot} \textbf{(c)}. Focus on the case where $\mu<0$ and $U>0$. When $U<\mu+1$, the ground state is the two-particle sector, and the sBZ Chern numbers are $\pm2$. When $U>\mu+3$ and $\mu>-1$, the ground state is doubly-degenerate and the states are in the single-particle sector. By adding a small spin-splitting term, we break the degeneracy and can calculate sBZ Chern numbers of $\pm1$. We now want to see what happens in the presence of a nonzero pairing.

To go beyond previous work~\cite{PhysRevB.105.184509} which studied both $s$- and $d$-wave pairings, we use a $p$-wave pairing which is natural in the context of topological superconductivity. Additionally, Ref. \cite{chan2025cherndartboardsuperconductors} argued that a $p$-wave pairing, in which different orbitals are paired, was necessary to realize non-trivial phases of an $n=2$ CDSC at the mean-field level. Consequently, we consider a pairing interaction of the form 
\begin{equation}
\label{eq:pairingterm}
H_{\text{pairing}}=-\frac{g}{L^{d}} \mathit{{\Delta}}^{\dagger}\mathit{{\Delta}},
\end{equation}
where $\mathit{{\Delta}} = \sum_{k} \Delta_{k}(c_{-k,1,\uparrow}c_{k,2,\downarrow} + c_{-k,2,\uparrow}c_{k,1,\downarrow})$, and $\Delta_{k}$ satisfies $\Delta_{-k} = -\Delta_{k}$. For a $p_x$-wave pairing, we set $\Delta_{k}=\sin(k_x)$, which preserves both mirror symmetries. At the mean-field level, the non-interacting CDSC has superconducting sBZ Chern numbers of $\pm4$ for a small nonzero $\Delta$.

As pointed out by Anderson and Haldane~\cite{ah}, Fermi liquids possess a hidden $Z_2$ symmetry.  The interaction in Eq.~\ref{hkint} is odd under the $Z_2$ symmetry and generates physics stemming from a fundamentally new fixed point~\cite{discrete, PWPZ2}. Moreover, the explicit $Z_2$ symmetry breaking must destroy a Fermi liquid. In fact, the essence of the $Z_2$ symmetry breaking is that it splits the single-pole in Fermi liquid theory into two poles. That is, the HK Green function consists of two poles corresponding to the lower and upper Hubbard bands at $\omega=\xi_{k}$ and $\omega=\xi_k+U$, respectively~\cite{phillips_exact_2020}.  For the purposes at hand, we need a 2-band generalization of this result. Consequently, we adopt the Green function
\beq
\label{eq:2BandGreenF}
G_{k\sigma}(\omega) &=& \sum_{m\in{\mathrm{I},\mathrm{II}}}\left(\frac{1- \braket{n_{m k\bar{\sigma}}}}{\omega-\xi_{m k}} + \frac{\braket{n_{m k\bar{\sigma}}}}{\omega-(\xi_{m k}+U)}\right),
\eeq
where $m$ sums over the bands in the CDI.  Consequently, interactions double the number of bands to four. We will refer to the new bands $\xi_{mk}$, $\xi_{mk} + U$ as the lower and upper Hubbard bands $\xi_{mk}^l$ and $\xi_{mk}^u$, namely $\rm LHB$ and $\rm UHB$ respectively, and they are shown in Fig.~\ref{fig:BandPlot} \textbf{(b)}. 


\subsection{Superconductivity}

\begin{figure}[ht]
    \centering
    \begin{minipage}[c]{0.9\linewidth}
        \centering
        \stackinset{l}{-3pt}{t}{-3pt}{\textbf{(a)}}{%
            \includegraphics[width=\linewidth]{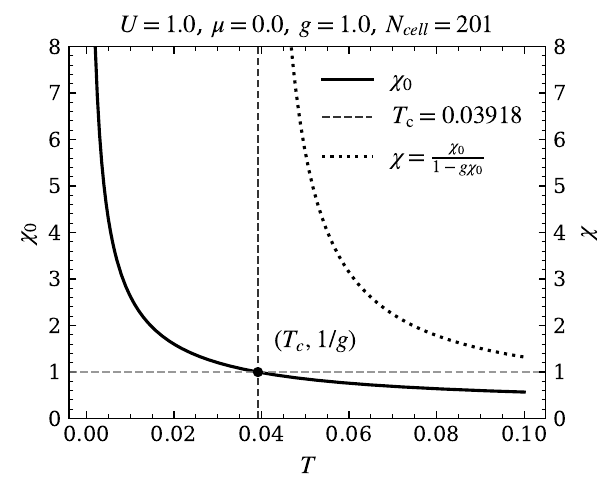}%
        }
        \label{fig:1(a)}
    \end{minipage}
    \par\vspace{0em}
    
    \begin{minipage}[c]{1.0\linewidth}
        \centering
        \stackinset{l}{5pt}{t}{0pt}{\textbf{(b)}}{%
            \includegraphics[width=\linewidth]{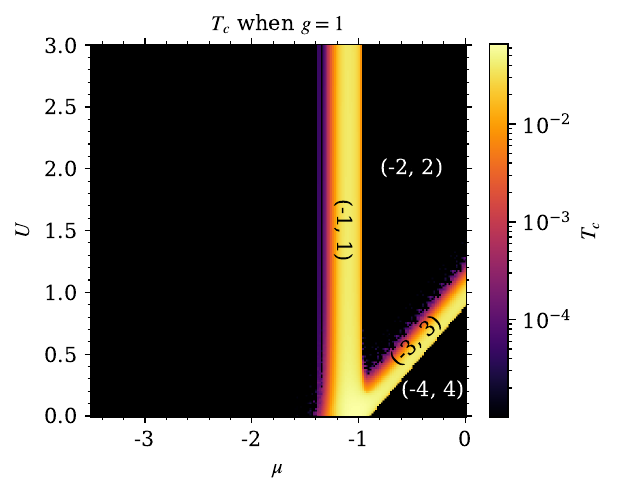}%
        }
        \label{fig:1(b)}
    \end{minipage}
    \caption{\textbf{(a)}~\ref{fig:1(a)} The bare and full pairing susceptibilities $\chi_0$, and $\chi$, respectively as a function of $T$. Here we chose $U = 1$ and $\mu = 0$, and set pairing strength $g = 1$. The divergence of $\chi_0$ as $T \rightarrow 0$ and $\chi$ as $T \rightarrow T_c$ confirm the superconducting phase.  \textbf{(b)}~\ref{fig:1(b)} The phase diagram of the critical temperature $T_c$ in our CDSC-HK model for $g=1$. Again the number pairs in different regions label the superconducting sBZ Chern numbers $(\mathcal{N}_r^a,\mathcal{N}_r^b)$ of our model.}
    \label{fig:PhaseDiagram_U_mu}
\end{figure}


\begin{figure*}[ht]
    \centering

    \begin{minipage}[c]{0.32\textwidth}
        \centering
        \stackinset{l}{3pt}{t}{3pt}{\textbf{(a)}}{%
            \includegraphics[width=\linewidth]{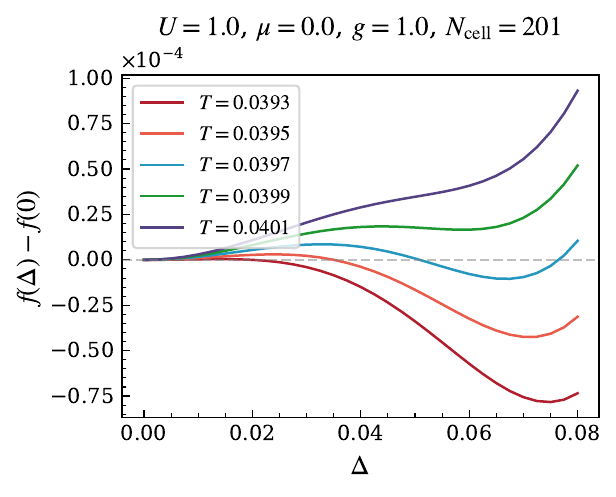}%
        }
    \end{minipage}
    \hfill
    \begin{minipage}[c]{0.32\textwidth}
        \centering
        \stackinset{l}{3pt}{t}{3pt}{\textbf{(b)}}{%
            \includegraphics[width=\linewidth]{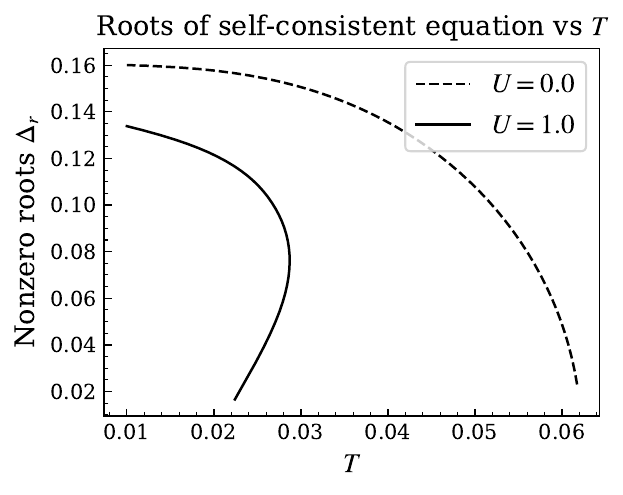}%
        }
    \end{minipage}
    \hfill
    \begin{minipage}[c]{0.32\textwidth}
        \centering
        \stackinset{l}{3pt}{t}{3pt}{\textbf{(c)}}{%
            \includegraphics[width=\linewidth]{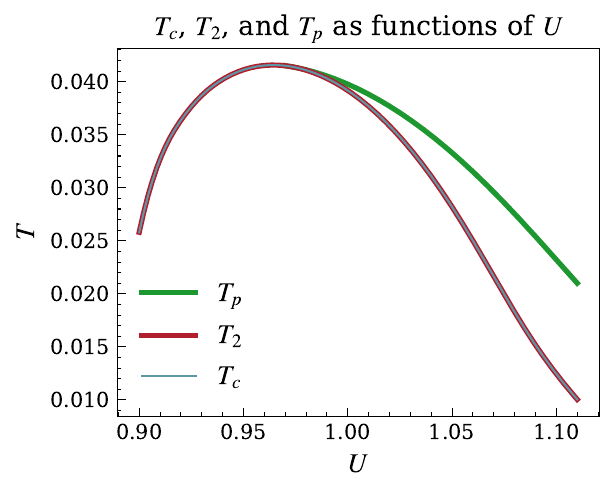}%
        }
    \end{minipage}
    
    \caption{\textbf{(a)} The shifted free energy, $f$, per unit volume, as a function of the average pairing amplitude $\overline{\Delta}$ at different temperatures (shifted by subtracting $f(0)$ for all the free energy curves to make sure they start from 0). The chemical potential $\mu$ is set to 0, the HK interaction strength $U = 1$, and the pairing strength $g = 1$. T varies from 0.033 to 0.042, and the system size is chosen as 201 by 201 cells. Clearly shown is the first-order nature for the turn-on of the pairing gap, $\Delta$. \textbf{(b)} The solution to the self-consistent equation as a function of $T$. The chemical potential, $\mu$ is the same as in (a). Though the momentum grid here is decreased to 21 by 21 for faster convergence, no finite-size effects were noticed. The dashed line corresponds to $U=0$ (the prior result of Chan/Hughes~\cite{chan2025cherndartboardsuperconductors}) and shows a monotonic decrease of the gap as the temperature increases. In contrast, for the interacting case (solid line), the behavior is non-monotonic, indicative of a first-order transition. \textbf{(c)} The variation of $T_c$, $T_2$, and $T_p$ as functions of $U$.  $T_c$ corresponds to the temperature at which the susceptibility diverges, $T_p$ the first-order transition temperature, and $T_2$ the temperature at which the free-energy has only a single minimum. To study the HK-induced superconducting phase, we chose $\mu = 0$ and constrain $U$ to $0.90 < U < 1.11$. As we can see, the $T_2$ and $T_c$ lines overlap, which means the inclined bright region yields a superconducting phase which follows the BCS dictum that $T_2=T_c$. The $T_p$ line signifies that this HK-induced superconductor can also have a first-order transition.}
    \label{fig:FreeEnergy_Delta&SelfConsistentSolution}
\end{figure*}

With this Green function in tow, we calculate the pairing susceptibility of our 2-band system. We formulate the bare pairing susceptibility  in imaginary time, $\tau$,  at $g = 0$ to be $\chi_0(\tau) = \braket{T_{\tau}\mathit{{\Delta}}(\tau)\mathit{{\Delta}}^{\dagger}(0)}_0$. Then the pairing susceptibility in terms of the Matsubara frequency $\nu_n$ is given by
\beq
	\chi_0(i\nu_n) = \frac{1}{L^d} \int_{0}^{\beta} d\tau e^{i\nu_n\tau}\chi_0(\tau).
\eeq
Through the Dyson equation $\chi = \chi_0 + g\chi_0\chi$, we can also obtain the full pairing susceptibility $\chi(i\nu_n)$ in terms of the Matsubara frequency.  As we show in the Appendix, the 2-band expression for the susceptibility reduces to
\begin{equation}
\begin{aligned}
    &\chi_0(i\nu_n) = \frac{1}{L^d} \sum_{a\in \{l, u\}} \sum_{k} |\Delta_k|^2 \bigg[ \\
    &(w_{\mathrm{I},k}^{11}w_{\mathrm{I},k}^{22} + w_{\mathrm{I},k}^{12}w_{\mathrm{I},k}^{21})(n_{\mathrm{I},k}^a)^2 \frac{2\tanh(\frac{\beta\xi_{\mathrm{I},k}^a}{2})}{2\xi_{\mathrm{I},k}^a - i\nu_n} \\
    +&(w_{\mathrm{II},k}^{11}w_{\mathrm{II},k}^{22} + w_{\mathrm{II},k}^{12}w_{\mathrm{II},k}^{21})(n_{\mathrm{II},k}^a)^2 \frac{2\tanh(\frac{\beta\xi_{\mathrm{II},k}^a}{2})}{2\xi_{\mathrm{II},k}^a - i\nu_n} \bigg]
\end{aligned}
\label{eq:chi0Simple}
\end{equation}
when $T \ll U$. We have defined the functions $w_{m,k}^{\alpha\beta}$ and $n_{m,k}^{a}$ in the Appendix and following sections as band spectral weight and state occupancy, respectively.

With the general 2-band formalism, and following the calculation in the Appendix, we arrive at our working equation
\begin{equation}
    \begin{aligned}
    &\chi_0(i\nu_n) = \frac{1}{L^d} \sum_{a\in \{l, u\}} \sum_{k} |\Delta_k|^2 \left(1 - \frac{d_z^2(\mathbf{k})}{E_k^2}\right) \\ 
    &\times\bigg[(n_{\mathrm{I},k}^a)^2 \frac{\tanh(\frac{\beta\xi_{\mathrm{I},k}^a}{2})}{2\xi_{\mathrm{I},k}^a - i\nu_n} + (n_{\mathrm{II},k}^a)^2 \frac{\tanh(\frac{\beta\xi_{\mathrm{II},k}^a}{2})}{2\xi_{\mathrm{II},k}^a - i\nu_n} \bigg]
\end{aligned}
\end{equation}
for the pairing susceptibility in the CDSC.

In Fig.~\ref{fig:PhaseDiagram_U_mu} \textbf{(a)}, we show the susceptibility as a function of temperature. As is evident, because the bare susceptibility diverges as $T$ is lowered, the full susceptibility must diverge as well when $g\chi_0=1$ at a temperature $T_c$.  For $U=1.0$, $\mu=0$, and $g=1.0$, we find a superconducting temperature of $T_c=0.03918$. Fig.~\ref{fig:PhaseDiagram_U_mu} \textbf{(b)} contains the full phase diagram when $g=1.0$ as a function of $U$ and band filling, $\mu$. All non-black regions indicate non-zero values of $T_c$. Two distinct superconducting regions stand out in this figure:
\begin{enumerate}
    \item The straight bright vertical region to the left of the $\mu = -1$ line, where $-1$ is the upper limit of the LHB for band $\mathrm{I}$.
    \label{feature:1}
    \item The slanted bright region in the vicinity of  $U = \mu + 1$ when $-1 < \mu < 0$.
    \label{feature:2}
\end{enumerate}
In region~\ref{feature:1}, SC persists to arbitrarily small and large $U$ because we always have a filling surface where the LHB of band $\mathrm{I}$ crosses the chemical potential. However, in region~\ref{feature:2} where $\mu>-1$, a finite HK interaction strength is required to generate SC. Interestingly, the repulsive interactions do not necessarily counteract superconductivity in this region, rather, they are an enabler. From the band structures shown in Fig.~\ref{fig:BandPlot}, we can understand this more clearly. When $-1 < \mu < 1$, the chemical potential is in the gap of the CDI as shown in Fig.~\ref{fig:BandPlot} \textbf{(a)}. Hence in this range, the BCS mechanism leads to a (non-diverging) saturation of the pair susceptibility, and without a strong enough pairing strength, even lowering the temperature to $T = 0$ does not lead to superconductivity. However, the HK interaction can still induce superconductivity by forcing the upper Hubbard bands to cross the chemical potential as shown in Fig.~\ref{fig:BandPlot} \textbf{(b)}. Such a crossing is sufficient to lead to a divergence of the pair susceptibility. In this sense, the HK interaction can induce superconductivity where the corresponding non-interacting system remains an insulator. Hence the  superconductivity here is driven by the repulsive interactions and, in this sense, is  non-traditional. Note, the pairing term considered here is identical in form to that used previously \cite{chan2025cherndartboardsuperconductors} and hence the nontrivial sBZ topology of the superconducting state remains intact. The HK interaction can induce new phases that have superconducting sBZ Chern numbers different from the non-interacting case, as shown in Fig.~\ref{fig:PhaseDiagram_U_mu} \textbf{(b)}. With no interaction, the SC sBZ Chern numbers are $\pm4$, twice the value of the sBZ Chern numbers of the insulator. However, by increasing $U$ and possibly changing $\mu$, we can obtain phases that have odd SC Chern numbers and which are not equivalent to a fine-tuned limit of an insulating phase.

\subsection{Thermodynamics}

\begin{figure*}[ht]
    \centering
    \begin{minipage}[c]{0.49\linewidth}
        \centering
        \stackinset{l}{0pt}{t}{0pt}{\textbf{(a)}}{%
            \includegraphics[width=\linewidth]{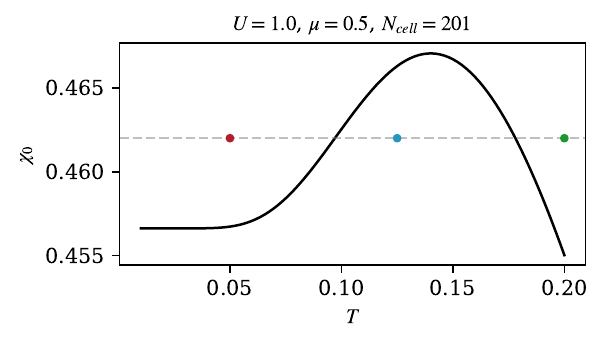}%
        }
    \end{minipage}
    \hfill
    \begin{minipage}[c]{0.49\linewidth}
        \centering
        \stackinset{l}{0pt}{t}{0pt}{\textbf{(b)}}{%
            \includegraphics[width=\linewidth]{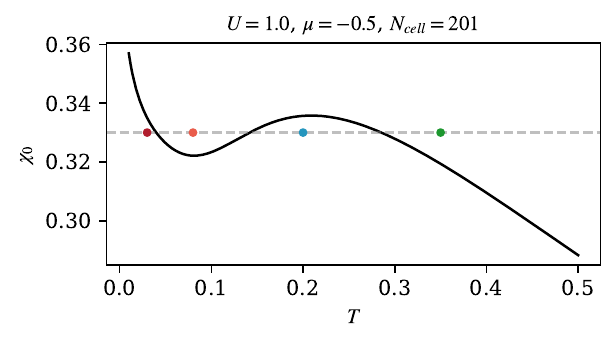}%
        }
    \end{minipage}

    \par\vspace{1em}

    \begin{minipage}[c]{0.49\linewidth}
        \centering
        \stackinset{l}{0pt}{t}{0pt}{\textbf{(c)}}{%
            \includegraphics[width=\linewidth]{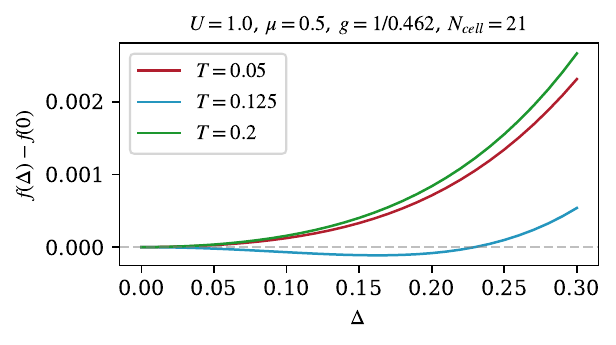}%
        }
    \end{minipage}
    \hfill
    \begin{minipage}[c]{0.49\linewidth}
        \centering
        \stackinset{l}{0pt}{t}{0pt}{\textbf{(d)}}{%
            \includegraphics[width=\linewidth]{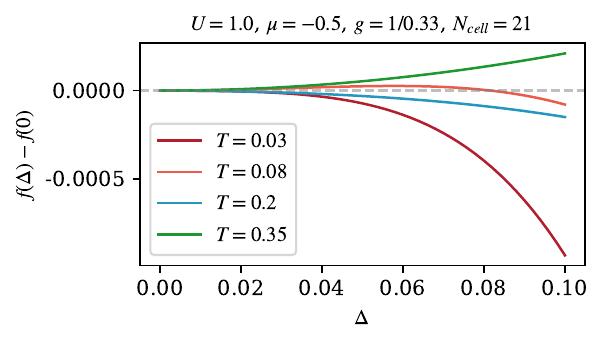}%
        }
    \end{minipage}
    \caption{\textbf{(a)}, \textbf{(c)} The behavior of $\chi_0(T)$ as a function of temperature for $\mu=0.5$ and $U = 1.0$. A peak in $\chi_0$ is followed by a subsequent decrease. The corresponding free-energy in \textbf{(c)} at the temperatures corresponds to the three dots shown in \textbf{(a)}, which illustrates that only the blue dot yields a superconductor. The green and red are insulating, hence there is an insulator-superconductor-insulator transition.
    \textbf{(b)}, \textbf{(d)} The behavior of $\chi_0(T)$ for $\mu=-0.5$ and $U = 1.0$. The peak in $\chi_0$ is followed by a subsequent decrease then increase again. The corresponding free-energy \textbf{(d)} at the temperatures corresponds to the four dots shown in \textbf{(b)}, which illustrates that the red and blue dots yield a superconductor. The green are insulating, while the orange indicates a state between the first and second phase transition, hence there is a more complicated re-entrant transition.}
    \label{fig:ReentrantChi0&FreeEnergy}
\end{figure*}

\begin{figure}
    \centering
    \includegraphics[width=1.0\linewidth]{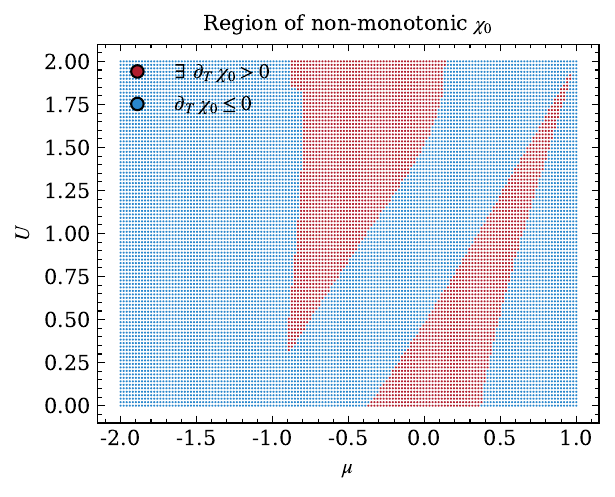}
    \caption{Phase diagram for non-monotonic $\chi_0$ of CDSC-HK system. The blue dots denote $\chi_0(T)$'s having first derivatives with respect to $T$ that are always $\leq$ 0 at those $(U, \mu)$ values, while the red dots denote $\chi_0(T)$'s where their first derivatives with respect to $T$ can be $>$ 0. Thus the red areas indicate $\chi_0$'s with non-monotonic behavior.}
    \label{fig:PhaseDiagramReentrant}
\end{figure}

The divergence of the susceptibility is sufficient to lay plain that the ground state is superconducting. Hence, with this result in hand we can perform a mean-field analysis to establish the turn-on of the gap. To facilitate this, we rely on the exact, finite-temperature formula for the state occupancy, $n_{\mathrm{I},k}^a$ and $n_{\mathrm{II},k}^a$, which is given by
\begin{equation}
    n_{\mathrm{I},k}^u = \frac{e^{-\beta\xi_{\mathrm{I},k}^{l}} + e^{-\beta(\xi_{\mathrm{I},k}^{l} + \xi_{\mathrm{I},k}^{u})}}{1 + 2 e^{-\beta\xi_{\mathrm{I},k}^{l}} + e^{-\beta(\xi_{\mathrm{I},k}^{l} + \xi_{\mathrm{I},k}^{u})}},
\end{equation}
where $n_{\mathrm{I},k}^l = 1 - n_{\mathrm{I},k}^u$. Similar formulas hold for both $n_{\mathrm{II},k}^l$ and $n_{\mathrm{II},k}^u,$ but with $\mathrm{I}$ replaced by $\mathrm{II}$. Following the 1-band result~\cite{PhysRevB.105.184509}, we apply a mean-field analysis on the $H_{\text{pairing}}$ term. Defining the pair amplitude $\Delta \equiv \frac{g}{L^d} \mathit{\Delta}$ and the deviation $\delta \Delta=\Delta-\overline{\Delta}$, we have that 
\begin{equation}
    \begin{aligned}
        H_{\text{pairing}}&=-\frac{g}{L^{d}}\mathit{{\Delta}}^{\dagger}\mathit{{\Delta}} = -\frac{L^{d}}{g}{\Delta}^{\dagger}{\Delta} \\
        &=-\frac{L^{d}}{g}(\delta\Delta^{\dagger} + \overline{\Delta}^*)(\delta\Delta + \overline{\Delta}) \\
        &\approx-\sum_k\big[\overline{\Delta}\Delta_{k}^*(c_{k,2,\downarrow}^{\dagger}c_{-k,1,\uparrow}^{\dagger} + c_{k,1,\downarrow}^{\dagger}c_{-k,2,\uparrow}^{\dagger}) \\
        & \quad \qquad+ \overline{\Delta}^*\Delta_{k}(c_{-k,1,\uparrow}c_{k,2,\downarrow} + c_{-k,2,\uparrow}c_{k,1,\downarrow})\big] \\
        & \quad \qquad+ \frac{L^d}{g}|\overline{\Delta}|^2 \\
        &= H_{\text{pairing}}^{\text{MF}} + \frac{L^d}{g}|\overline{\Delta}|^2,
    \end{aligned}
\end{equation}
where $\overline{\Delta} \equiv \braket{\Delta}$ is the average with respect to the thermal density matrix.
Finally, our total Hamiltonian under mean-field theory becomes 
\begin{equation}
    \begin{aligned}
        H &= H_0 + H_{\text{HK}} + H_{\text{pairing}}^{\text{MF}} + \frac{L^d}{g}|\overline{\Delta}|^2 \\
        &\equiv \sum_{k \in \text{HFBZ}} H_{k}^{\text{MF}} + \frac{2L^d}{g}|\overline{\Delta}|^2,
    \end{aligned}
\end{equation}
which is block diagonalized in $\textbf{k}$ sectors~\cite{TopoSWaveSC}, and the summation over $\textbf{k}$ is performed on half of the first Brillouin zone (HFBZ) to avoid double counting, hence the factor of 2 in the numerator in the last term.  

The self-consistent equation and free energy $F$ for this model are given by
\begin{equation}
    \overline{\Delta} = \frac{g}{L^d}\sum_k\Delta_{k}\braket{c_{-k,1,\uparrow}c_{k,2,\downarrow} + c_{-k,2,\uparrow}c_{k,1,\downarrow}},
\end{equation}
and
\begin{equation}
    F(\overline{\Delta}) = -\frac{1}{\beta}\ln{Z} + \frac{2L^d}{g}|\overline{\Delta}|^2,
\end{equation}
with
\begin{equation}
    Z = \prod_{k \in \text{HFBZ}}\text{Tr}(e^{-\beta H_{k}^{\text{MF}}}).
\end{equation}

Focusing on the HK-induced superconductor in our model, we plot the graph for the free energy per volume $f = F/L^d$ versus the average pairing amplitude $\overline{\Delta}$ at different temperatures. Abundantly clear from Fig.~\ref{fig:FreeEnergy_Delta&SelfConsistentSolution} \textbf{(a)} is that at high enough temperature, $\Delta=0$ is the only minimum. At sufficiently low temperatures, a new minimum appears with $\Delta \ne 0$. This is mapped out in Fig.~\ref{fig:FreeEnergy_Delta&SelfConsistentSolution} \textbf{(b)} by the solid line.  At $U=0$, the two-solution region vanishes giving rise to a pure second-order transition. From the free energy, we can define $T_2$ to be the highest temperature at which a single $\Delta \ne 0$ minimum exists. We also define $T_p$ as the temperature at which the free energy exhibits two degenerate minima. In BCS, $T_2=T_p=T_c$, which is not true in general for the strongly interacting case. The behavior of these three temperature scales is clearly shown in Fig.~\ref{fig:FreeEnergy_Delta&SelfConsistentSolution} \textbf{(c)}.
Similar results have been obtained previously ~\cite{PhysRevB.105.184509, TwoStageHKBCSSC, hahn2026multibandsuperconductivityexactlysolvable}.


\subsection{Re-entrance Phenomena}

We now point out a feature unique to the onset of superconductivity in the CDI platform.  Namely, we find a re-entrant insulating phase which is absent from the 1-band HK SC system. For BCS theory in a metallic 1-band system with $s$-wave pairing, the bare pairing susceptibility $\chi_0$ will increase monotonically as $T$ decreases. For an insulating phase, however, $\chi_0$ saturates monotonically as the temperature decreases. Regardless of the parameters used,  as long as  $g$ intersects with $\chi_0$, we will obtain a critical temperature $T_c$. In the CDSC, a third possibility is apparent from Fig.~\ref{fig:ReentrantChi0&FreeEnergy} \textbf{(a)}. We see that the initial rise in the susceptibility as $T$ decreases is accompanied by a subsequent decrease that will ultimately reach a constant value. Fig.~\ref{fig:ReentrantChi0&FreeEnergy} \textbf{(c)} shows the corresponding free energy at three chosen points, and we see that the middle temperature constitutes the only superconducting phase. In contrast, both high- and low-temperature regimes exhibit insulating phases.
More precisely, the reason our system supports re-entrance is that $\chi_0$ reaches a peak at a finite temperature (named $T_{peak}$), then decreases and then converges to a lower value as $T$ decreases. Thus, we can take a $g$ value (which we refer to as $g_0$) such that it satisfies $\chi_0(T = 0) < 1/g_0 < \chi_0(T = T_{peak})$, and the $1/g_0$ line will intersect with the $\chi_0$ curve at 2 points; we call the temperatures corresponding to these 2 points as $T_i$ and $T_f$. When $T < T_i$, $\chi_0 < 1/g_0$, so the system is in a normal phase. When $T_i < T < T_f$, $\chi_0 > 1/g_0$, and the system is in the superconducting phase. When $T_f < T$, $\chi_0 < 1/g_0$ again, so the system enters normal phase again.
Moreover, Fig.~\ref{fig:ReentrantChi0&FreeEnergy} \textbf{(b)} and \textbf{(d)} show an even more complicated situation. Here $\chi_0$ goes through a process of decreasing, increasing, and then decreasing again. The corresponding free energy at the orange point shows a state between the first and the second order transitions. Finally, we can obtain a SC-SC with metastable metal-SC-metal transition.

This kind of non-monotonic behavior of $\chi_0$ is not rare in our CDSC-HK system. Fig.~\ref{fig:PhaseDiagramReentrant} plots the phase diagram of our system in the $(U, \mu)$ plane where non-monotonicity in $\chi_0$ obtains. We use the first derivative of $\chi_0$ on $T$ to detect the non-monotonic behavior, demarcated by the red regions in Fig. ~\ref{fig:PhaseDiagramReentrant}.
From where does this non-monotonic behavior arise? Compared with a single band superconductor without HK interactions, the two different ingredients in our model are as follows:
\begin{enumerate}
    \item The HK interaction creates two poles in the Green function, each has a temperature dependent weight given by state occupancy. Thus, the pairing susceptibility contains terms $\chi_{0,mm'}^{lu}$ and $\chi_{0,mm'}^{ul}$ $(m,m'\in\{\mathrm{I},\mathrm{II}\})$ that mix the two poles, the $T$-derivatives of which are complicated.
    \label{ingredient:1}
    \item The inter-orbital pairing creates inter-band terms $\chi_{0,\mathrm{I}\,\mathrm{II}}$ and $\chi_{0,\mathrm{II}\,\mathrm{I}}$ in the pairing susceptibility, whose derivatives with respect to $T$ are also not simple.
    \label{ingredient:2}
\end{enumerate}
Both of these factors can contribute to the non-monotonic behavior of $\chi_0(T)$, and thus lead to a non-trivial re-entrant behavior. In the Appendix, we perform the detailed temperature derivatives of all of these terms, comparing with the temperature derivative of a simple single band superconductivity without HK to show their difference analytically.


To our knowledge, temperature-driven re-entrant phenomena in superconductivity has previously been discussed in Kondo systems~\cite{KondoEffectinSuperconductors, SecondTransitionKondoEffect}, population-imbalanced Fermi gases~\cite{AtomicFermiGaswithPopulationImbalance}, hybridized Anderson lattices~\cite{vanGervenOei_2020}, magnetic proximity structures~\cite{ConicalFerromagnetSuperconductorNanostructures}, and Josephson arrays~\cite{avraham_reentrant_2026}. However, reentrance originating from a non-monotonic bare pairing susceptibility generated by the HK interactions and inter-orbital pairings has not been explicitly identified. Such findings and discussions above undoubtedly provide a new approach to realize re-entrant phenomena in superconductivity.




\section{Conclusion}

In this work, we have systematically extended the 1-band HK superconductor to a 2-band setting by introducing an exact theory of the pairing susceptibility.  Our numerical and mean-field results for a specific Chern dartboard insulator model lead us to identify a new kind of superconducting phase induced by HK interactions that has non-trivial sub-Brillouin zone Chern numbers. Moreover, from the self-consistent solutions, we were able to identify that inter-orbital pairing leads to a unique zero-frequency low temperature insulator-superconductor-insulator re-entrant phase transition. 

That 2-band superconductivity with HK interactions contains re-entrance is an example of `more' being quite different.  That is, the re-entrance, not present in either of the bands separately, emerges truly from the 2-band complexity.  Looking ahead, this 2-band inter-orbital pairing HK superconductor raises many possible questions. An immediate next step is to test the robustness of the HK-induced superconducting phase and the re-entrant phenomenon by changing the $H_0$ in the Hamiltonian to other models, or changing the band HK interaction to those of the momentum-mixing kind.  More generally, our exact theory could be applied to other superconducting systems with different pairing forms, or to systems with different spatial dimensions, which may potentially unveil richer topological and/or re-entrant superconducting phases.

\textbf{Acknowledgement}  P.W.P. acknowledges funding from the Research Board of the University of Illinois, CRB Award RB26125, for partial funding of this project. R.C. and T.L.H. thank ARO MURI W911NF2020166 for support. R.C. thanks Gaurav Tenkila for useful discussions on HK.
\newline

\vfill

\bibliographystyle{apsrev4-2}
\bibliography{my_ref}

\clearpage

\widetext
\begin{center}
\textbf{\large Appendix: Derivation of Bare Pairing Susceptibility for the 2-Band Superconductor Model}
\end{center}
\setcounter{equation}{0}
\setcounter{figure}{0}
\setcounter{table}{0}
\setcounter{page}{1}
\makeatletter
\renewcommand{\theequation}{S\arabic{equation}}
\renewcommand{\thefigure}{S\arabic{figure}}
\renewcommand{\bibnumfmt}[1]{[S#1]}
\renewcommand{\citenumfont}[1]{S#1}
\makeatother

In this Supplementary Material, we present detailed derivations of $\chi_0(i\nu_n)$ and $\chi_0(q; i\nu_n)$ on our specific Chern dartboard insulator with inter-orbital pairing and band HK interaction.

\section{S-I. Derivation of $\chi_0(i\nu_n)$ and $\chi_0(q; i\nu_n)$}

Focusing on $\chi_0(\tau) = \braket{T_{\tau}\mathit{{\Delta}}(\tau)\mathit{{\Delta}}^{\dagger}(0)}_0$, substituting our inter-orbital $p$-wave pairing term,  and simplifying the equation,
\begin{equation}
\begin{aligned}
    \chi_0(\tau) = \sum_{k} |\Delta_k&|^2[G_{0,-k,11,\uparrow}(\tau) G_{0,k,22,\downarrow}(\tau) \\ &+G_{0,-k,22,\uparrow}(\tau) G_{0,k,11,\downarrow}(\tau)\\
    \\ &+G_{0,-k,12,\uparrow}(\tau) G_{0,k,21,\downarrow}(\tau)\\
    \\ &+G_{0,-k,21,\uparrow}(\tau) G_{0,k,12,\downarrow}(\tau)],
    \label{chi0}
\end{aligned}
\end{equation}
we obtain a diagonal equation in terms of the $k,-k$ states.  These are the only states mixed by our starting Hamiltonian.
For a non-interacting system,
\begin{equation}
    G_{0,k,\alpha\beta,\sigma}(\tau) = \sum_n e^{-i\nu_n\tau} [i\nu_n-H_0(k)]^{-1}_{\alpha\beta,\sigma}
\end{equation}
is the operative equation for the Green function.
Here, $\alpha,\beta \in \{1, 2\}$ are the orbital indices.
However, with the HK interaction, we need to consider the HK "lifted" part for each Green function. Upon simplification, we finally obtain the following formula for the Green functions used above:
\begin{equation}
\begin{aligned}
    &-G_{0,k,\alpha\beta,\sigma}(\tau) \\
    &= w_{\mathrm{I},k}^{\alpha\beta}\big[(1-\braket{n_{\mathrm{I},k,\bar\sigma}}) f(-\xi_{\mathrm{I},k})e^{-\xi_{\mathrm{I},k} \tau}\\
    &\qquad \quad+ \braket{n_{\mathrm{I},k,\bar\sigma}} f(-(\xi_{\mathrm{I},k} + U))e^{-\tau(\xi_{\mathrm{I},k} + U)} \big]\\
    &+ w_{\mathrm{II},k}^{\alpha\beta}\big[(1-\braket{n_{\mathrm{II},k,\bar\sigma}}) f(-\xi_{\mathrm{II},k})e^{-\xi_{\mathrm{II},k} \tau} \\
    &\qquad \quad+ \braket{n_{\mathrm{II},k,\bar\sigma}} f(-(\xi_{\mathrm{II},k} + U))e^{-\tau(\xi_{\mathrm{II},k} + U)} \big].
\end{aligned}
\end{equation}
Here, $w_{\mathrm{I},k}^{\alpha\beta}$ and $w_{\mathrm{II},k}^{\alpha\beta}$ are the band spectral weights and are defined as the elements of the band projector matrix: $w_{m, k}^{\alpha\beta} = [P_{m,k}]_{\alpha\beta}$ for $m \in \{\mathrm{I}, \mathrm{II}\}$.
Specifically in our model, $P_{\mathrm{I},k} = 1/2[I-H_2/E_k]$, and $P_{\mathrm{II},k} = 1/2[I+H_2/E_k]$. Thus the spectral weights are given by
\begin{equation}
\begin{aligned}
    &w_{\mathrm{I},k}^{22} = w_{\mathrm{II},k}^{11} = \frac{1}{2}(1 + \frac{d_z(\mathbf{k})}{E_k}), \\
    &w_{\mathrm{I},k}^{11} = w_{\mathrm{II},k}^{22} = \frac{1}{2}(1 - \frac{d_z(\mathbf{k})}{E_k}), \\
    &w_{\mathrm{II},k}^{12} = -w_{\mathrm{I},k}^{12} = \frac{1}{2}(\frac{d_x(\mathbf{k})-id_y(\mathbf{k})}{E_k}), \\
    &w_{\mathrm{II},k}^{21} = -w_{\mathrm{I},k}^{21} = \frac{1}{2}(\frac{d_x(\mathbf{k})+id_y(\mathbf{k})}{E_k}).
\end{aligned}
\end{equation}


Substitution of these expressions into Eq. (\ref{chi0}) results in our working expression
\begin{equation}
\begin{aligned}
    &\chi_0(i\nu_n) = \frac{1}{L^d} \sum_{a,b \in \{l, u\}} \sum_{k} |\Delta_k|^2 \bigg[ \\
    &(w_{\mathrm{I},k}^{11}w_{\mathrm{I},k}^{22} + w_{\mathrm{I},k}^{12}w_{\mathrm{I},k}^{21})n_{\mathrm{I},k}^a n_{\mathrm{I},k}^b \frac{\tanh(\frac{\beta\xi_{\mathrm{I},k}^a}{2})+\tanh(\frac{\beta\xi_{\mathrm{I},k}^b}{2})}{\xi_{\mathrm{I},k}^a + \xi_{\mathrm{I},k}^b - i\nu_n} \\
    +&(w_{\mathrm{II},k}^{11}w_{\mathrm{II},k}^{22} + w_{\mathrm{II},k}^{12}w_{\mathrm{II},k}^{21})n_{\mathrm{II},k}^a n_{\mathrm{II},k}^b \frac{\tanh(\frac{\beta\xi_{\mathrm{II},k}^a}{2})+\tanh(\frac{\beta\xi_{\mathrm{II},k}^b}{2})}{\xi_{\mathrm{II},k}^a + \xi_{\mathrm{II},k}^b - i\nu_n} \\
    +&(w_{\mathrm{I},k}^{11}w_{\mathrm{II},k}^{22} + w_{\mathrm{I},k}^{12}w_{\mathrm{II},k}^{21})n_{\mathrm{I},k}^a n_{\mathrm{II},k}^b \frac{\tanh(\frac{\beta\xi_{\mathrm{I},k}^a}{2})+\tanh(\frac{\beta\xi_{\mathrm{II},k}^b}{2})}{\xi_{\mathrm{I},k}^a + \xi_{\mathrm{II},k}^b - i\nu_n} \\
    +&(w_{\mathrm{II},k}^{11}w_{\mathrm{I},k}^{22} + w_{\mathrm{II},k}^{12}w_{\mathrm{I},k}^{21})n_{\mathrm{II},k}^a n_{\mathrm{I},k}^b \frac{\tanh(\frac{\beta\xi_{\mathrm{II},k}^a}{2})+\tanh(\frac{\beta\xi_{\mathrm{I},k}^b}{2})}{\xi_{\mathrm{II},k}^a + \xi_{\mathrm{I},k}^b - i\nu_n}\bigg]
\end{aligned}
\end{equation}
for the pair susceptibility.
To clarify these expressions,  the numbers $1$ and $2$ indicate the 2 orbitals, $\mathrm{I}$ and $\mathrm{II}$ in the subscripts indicate the 2 bands, and the superscripts $a,b \in \{l, u\}$ represent indicate the lower and upper bands in the HK model.  
As we can see, the pairing susceptibility above can be divided into 4 main parts, namely
\begin{equation}
    \chi_0 = \chi_{0,\mathrm{I}\,\mathrm{I}} + \chi_{0,\mathrm{II}\,\mathrm{II}} + \chi_{0,\mathrm{I}\,\mathrm{II}} + \chi_{0,\mathrm{II}\,\mathrm{I}},
    \label{eq:chi0subterms}
\end{equation}
and each part has 4 sub-parts that come from the sum of $a,b \in \{l, u\}$, for example, $\chi_{0,\mathrm{I}\,\mathrm{I}} = \chi_{0,\mathrm{I}\,\mathrm{I}}^{ll} + \chi_{0,\mathrm{I}\,\mathrm{I}}^{lu} + \chi_{0,\mathrm{I}\,\mathrm{I}}^{ul} + \chi_{0,\mathrm{I}\,\mathrm{I}}^{uu}$.
When $T \ll U$, the contribution of $\chi_{0,\mathrm{I}\,\mathrm{II}} + \chi_{0,\mathrm{II}\,\mathrm{I}}$ is negligibly small, and the sub-parts of $a \neq b$ are also minimal. Thus we can simplify $\chi_0(i\nu_n)$ in Eq.~\ref{eq:chi0Simple} by dropping the terms involving the upper Hubbard bands.

Similarly, for the center-of-mass momentum $q \ne 0$ case, $\chi_0(q;i\nu_n)$ takes on the form
\begin{equation}
\begin{aligned}
    &\chi_0(q;i\nu_n) = \frac{1}{L^d} \sum_{a,b \in \{l, u\}} \sum_{k} |\Delta_k|^2 \bigg[ \\
    & w_{\mathrm{I},k;\mathrm{I},k+q}^{11\cdot22+12\cdot21} n_{\mathrm{I},k}^a n_{\mathrm{I},k+q}^b \frac{\tanh(\frac{\beta\xi_{\mathrm{I},k}^a}{2})+\tanh(\frac{\beta\xi_{\mathrm{I},k+q}^b}{2})}{\xi_{\mathrm{I},k}^a + \xi_{\mathrm{I},k+q}^b - i\nu_n} \\
    +& w_{\mathrm{II},k;\mathrm{II},k+q}^{11\cdot22+12\cdot21} n_{\mathrm{II},k}^a n_{\mathrm{II},k+q}^b \frac{\tanh(\frac{\beta\xi_{\mathrm{II},k}^a}{2})+\tanh(\frac{\beta\xi_{\mathrm{II},k+q}^b}{2})}{\xi_{\mathrm{II},k}^a + \xi_{\mathrm{II},k+q}^b - i\nu_n} \\
    +& w_{\mathrm{I},k;\mathrm{II},k+q}^{11\cdot22+12\cdot21} n_{\mathrm{I},k}^a n_{\mathrm{II},k+q}^b \frac{\tanh(\frac{\beta\xi_{\mathrm{I},k}^a}{2})+\tanh(\frac{\beta\xi_{\mathrm{II},k+q}^b}{2})}{\xi_{\mathrm{I},k}^a + \xi_{\mathrm{II},k+q}^b - i\nu_n} \\
    +& w_{\mathrm{II},k;\mathrm{I},k+q}^{11\cdot22+12\cdot21} n_{\mathrm{II},k}^a n_{\mathrm{I},k+q}^b \frac{\tanh(\frac{\beta\xi_{\mathrm{II},k}^a}{2})+\tanh(\frac{\beta\xi_{\mathrm{I},k+q}^b}{2})}{\xi_{\mathrm{II},k}^a + \xi_{\mathrm{I},k+q}^b - i\nu_n} \\
    &\bigg].
\end{aligned}
\end{equation}

\clearpage

\section{S-II. First Derivative of $\chi_0(T)$ on Temperature $T$}

Following Eq.~\ref{eq:chi0subterms}, $\chi_0$ has the component $\chi_{0,\mathrm{I}\,\mathrm{I}} + \chi_{0,\mathrm{II}\,\mathrm{II}}$ contributed by intra-band pairing, and the term $\chi_{0,\mathrm{I}\,\mathrm{II}} + \chi_{0,\mathrm{II}\,\mathrm{I}}$ arising from inter-band pairing. Also, each $\chi_{0,mm'}$ ($m, m' \in \{\mathrm{I}, \mathrm{II}\}$) term is composed of $\chi_{0,mm'}^{ll}$, $\chi_{0,mm'}^{lu}$, $\chi_{0,mm'}^{ul}$, and $\chi_{0,mm'}^{uu}$. The term $\chi_{0,mm'}^{lu} + \chi_{0,mm'}^{ul}$ mixed the two poles provided by the HK interaction, while $\chi_{0,mm'}^{ll} + \chi_{0,mm'}^{uu}$ remain unchanged. Following this, we find that our first derivative of $\chi_0$ with respect to temperature $T$ is composed of 2 parts, namely, intra-band term defined as $\sum_{m\in\{\mathrm{I}, \mathrm{II}\};a,b\in\{l,u\}}\partial_T(\chi_{0,mm}^{ab})$ and inter-band term defined as $\sum_{a,b\in\{l,u\}}\partial_T(\chi_{0,\mathrm{I}\,\mathrm{II}}^{ab})$.
For intra-band term we have
\begin{align}
    &\partial_T(\chi_{0,mm}^{ab}) = \partial_T\sum_{k} |\Delta_k|^2 (1 - \frac{d_z^2(\mathbf{k})}{E_k^2})n_{m,k}^a n_{m,k}^b \frac{\tanh(\frac{\beta\xi_{m,k}^a}{2}) + \tanh(\frac{\beta\xi_{m,k}^b}{2})}{\xi_{m,k}^a + \xi_{m,k}^b} \nonumber \\
    &=\sum_{k} |\Delta_k|^2 (1 - \frac{d_z^2(\mathbf{k})}{E_k^2})\Big[
    \partial_T(n_{m,k}^a n_{m,k}^b)\frac{\tanh(\frac{\beta\xi_{m,k}^a}{2}) + \tanh(\frac{\beta\xi_{m,k}^b}{2})}{\xi_{m,k}^a + \xi_{m,k}^b} -n_{m,k}^a n_{m,k}^b\frac{\xi_{m,k}^a\text{sech}^2(\frac{\beta\xi_{m,k}^a}{2}) + \xi_{m,k}^b\text{sech}^2(\frac{\beta\xi_{m,k}^b}{2})}{2T^2(\xi_{m,k}^a + \xi_{m,k}^b)} \Big].
\end{align}
For the inter-band term, we have
\begin{align}
    &\partial_T(\chi_{0,\mathrm{I}\,\mathrm{II}}^{ab}) = \partial_T\sum_{k} |\Delta_k|^2 (\frac{d_z^2(\mathbf{k})}{E_k^2}) n_{\mathrm{I},k}^a n_{\mathrm{II},k}^b \frac{\tanh(\frac{\beta\xi_{\mathrm{I},k}^a}{2}) + \tanh(\frac{\beta\xi_{\mathrm{II},k}^b}{2})}{\xi_{\mathrm{I},k}^a + \xi_{\mathrm{II},k}^b} \nonumber \\
    &=\sum_{k} |\Delta_k|^2 (\frac{d_z^2(\mathbf{k})}{E_k^2})\Big[\partial_T(n_{\mathrm{I},k}^a n_{\mathrm{II},k}^b)\frac{\tanh(\frac{\beta\xi_{\mathrm{I},k}^a}{2}) + \tanh(\frac{\beta\xi_{\mathrm{II},k}^b}{2})}{\xi_{\mathrm{I},k}^a + \xi_{\mathrm{II},k}^b} - n_{\mathrm{I},k}^a n_{\mathrm{II},k}^b\frac{\xi_{\mathrm{I},k}^a\text{sech}^2(\frac{\beta\xi_{\mathrm{I},k}^a}{2}) + \xi_{\mathrm{II},k}^b\text{sech}^2(\frac{\beta\xi_{\mathrm{II},k}^b}{2})}{2T^2(\xi_{\mathrm{I},k}^a + \xi_{\mathrm{II},k}^b)} \Big].
\end{align}

Thus we can see, both the temperature dependent state occupancy for different poles created by HK interaction and the inter-band pairing contribute to the non-monotonic behavior of $\chi_0(T)$, and are the efficient cause of the re-entrant phenomena. Incomparison, we find that a single band superconductor without HK interactions has a $\chi_0$ and its derivative on $T$ as
\begin{align}
    \partial_T(\chi_0) &= \sum_{k} |\Delta_k|^2 \partial_T \frac{\tanh(\frac{\beta\xi_{k}}{2})}{2\xi_{k}} = -\sum_{k} |\Delta_k|^2 \frac{1}{4T^2}\text{sech}^2(\frac{\beta\xi_{k}}{2}),
\end{align}
which is strictly $< 0$ and hence no re-entrance.

\end{document}